\documentclass[final]{aipproc}
\layoutstyle{6x9}

\usepackage{amsmath}
\usepackage{multirow}

\newcommand{\bra}[1]{\langle #1|}
\newcommand{\ket}[1]{|#1\rangle}
\newcommand{\braket}[1]{\langle #1\rangle}
\newcommand{\Tr}[1]{\mathrm{Tr}\left[#1\right]}

\begin{document}

\title{Hadronic Resonances from Lattice QCD}

\author{John Bulava}{address={Department of Physics, Carnegie Mellon University, Pittsburgh, PA 15213, USA}}
\author{Robert Edwards}{address={Thomas Jefferson National Accelerator 
Facility, Newport News, VA 23606, USA}}
\author{George Fleming}{address={Yale University, New Haven, CT 06520, USA}}
\author{K. Jimmy Juge}{address={Department of Physics, University of the Pacific, Stockton, CA 95211, USA}}
\author{Adam C. Lichtl\footnote{Speaker}\space}{address={RBRC, Brookhaven National Laboratory, Upton, NY 11973,USA}}
\author{Nilmani Mathur}{address={Thomas Jefferson National Accelerator Facility, Newport News, VA 23606, USA}} 
\author{Colin Morningstar}{address={Department of Physics, Carnegie Mellon University, Pittsburgh, PA 15213, USA}}
\author{David Richards}{address={Thomas Jefferson National Accelerator Facility, Newport News, VA 23606, USA}}
\author{Stephen J. Wallace}{address={University of Maryland, College Park, MD 20742, USA}}

\begin{abstract}
The determination of the pattern of hadronic resonances as predicted by Quantum Chromodynamics requires the use of non-perturbative techniques.  Lattice QCD has emerged as the dominant tool for such calculations, and has produced many QCD predictions which can be directly compared to experiment.  The concepts underlying lattice QCD are outlined, methods for calculating excited states are discussed, and results from an exploratory Nucleon and Delta baryon spectrum study are presented.
\end{abstract}

\maketitle

\section{Motivation}
Patterns observed in measured spectra have repeatedly inspired fundamental breakthroughs in modeling reality.  The periodic table of the chemical elements led to the construction of the atomic model, and atomic spectroscopy led to the development of quantum mechanics.   Later, the categorization of subatomic particles into the geometric patterns of the `Eightfold Way' spawned the creation of the quark model and ultimately, the advent of Quantum Chromodynamics (QCD).

It is critical to the development of particle physics to determine if QCD gives rise to experimentally observed nuclear physics phenomena.  The QCD spectrum is a particularly desirable calculation not only because it would provide a check of the theory, but also because it could lead to a much deeper understanding of the physics of the strong interaction.  Additionally, because its underlying quark degrees of freedom are color confined, QCD's low-energy behavior must be inferred from the properties of its composite hadronic resonances.  

One of the main goals of the Lattice Hadron Physics Collaboration (LHPC) is to calculate the QCD hadron spectrum from first principles using the lattice QCD formalism.  Such a calculation would shed light on many interesting phenomenological issues such as the nature of the Roper resonance, the missing quark model resonances, and the properties of hybrid and exotic hadronic resonances.

\section{Lattice QCD as a tool to study resonances}

\subsection{The need for the lattice formalism}
In QCD the effective strength of the strong nuclear force is controlled by the renormalized coupling between the quarks and the gluons.  This quantity is dependent on the energy scale of the interaction and becomes weak at high-energies, leading to the property of asymptotic freedom.  This is consistent with the results of deep inelastic scattering experiments involving high momentum transfer interactions.  At these scales, the application of perturbation theory has led to predictions which agree spectacularly with experiment.

In contrast, the coupling becomes strong at lower energy scales.  This is consistent with color confinement, and leads to a rich spectral structure.  This strong coupling precludes any perturbative expansion (in the coupling) for low energy quantities such as the spectrum.  Fortunately, non-perturbative calculations are possible using a space-time lattice regulator in conjunction with the functional integral framework~\cite{bib:wilson}.  The use of a lattice not only defines a measure for the functional integral, but also introduces an ultraviolet cutoff via the inverse lattice spacing $a^{-1}$.  In the following we survey several key concepts which enable lattice practitioners to calculate the QCD spectrum.

\subsection{The physics of lattice QCD}
\begin{figure}[ht!]
  \includegraphics[height=0.2\textheight]{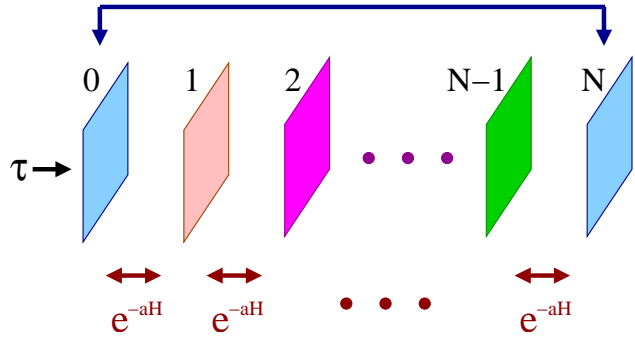}
  \caption{A graphical representation of the system lattice QCD simulates: $N$ time slices of quantum field variables coupled by the transfer matrix $T=\exp(-aH)$.  Periodic temporal boundary conditions identify the fields on the first ($\tau = 0$) time slice with those on the the last ($\tau=Na$) time slice.\label{fig:lattice}}
\end{figure}

Several expository books on lattice QCD are available~\cite{bib:books}.  Here we review those features of the formalism relevant to spectroscopy, using scalar fields for simplicity.  In the continuum, quantum fields $\phi(\vec{x}; t)$ are defined over all spatial points $\vec{x}$ at each time $t$, and are related at different times through the {\em time evolution operator}:
\begin{equation}
\exp[-iH(t-t')],
\end{equation} where $H$ is the Hamiltonian of the theory.  The lattice formulation uses fields restricted to the sites of a $4$-dimensional hypercubic lattice.  The lattice fields are denoted $\phi_\tau(\vec{x})$, where $\vec{x}$ is the spatial index now restricted to values on a $3$-dimensional cubic lattice.  The discrete index $\tau$ does not represent the standard (Minkowski) time $t$, but instead denotes imaginary (Euclidean) time.  The use of Euclidean time implies that the collection of fields on one time slice is related to the collection of fields on a neighboring time slice via the {\em transfer matrix}:
\begin{equation}
T=\exp[-aH],
\end{equation}
where $a$ is the lattice spacing between neighboring time slices.
Using the exponential Euclidean transfer matrix allows us to utilize many computational techniques from statistical mechanics not available when using the oscillatory Minkowski time evolution operator.
 If we use a finite temporal extent with periodic boundary conditions, we have the system shown in Fig.~\ref{fig:lattice}.  

If we perform an integral over the possible values of the field variables on all of the time slices $\{\phi_\tau\}$, we get the partition function of the theory:
\begin{eqnarray}
Z&=&\int\,\mathcal{D}\phi\,\braket{\phi_0 | e^{-a H} | \phi_{N-1}}\cdots\braket{\phi_1 | e^{-a H} | \phi_0}, \quad \mathcal{D}\phi \equiv \Pi_\tau d\phi_\tau,\label{eqn:partition1}\\
&=&\int\,d\phi_0\,\braket{\phi_0 | e^{-\beta H} | \phi_0}=\Tr{e^{-\beta H}}, \quad \beta \equiv N_\tau a,\label{eqn:partition2}
\end{eqnarray}
where we have used the completeness relation
$\int\,d\phi_\tau \ket{\phi_\tau}\bra{\phi_\tau} = 1.$

\paragraph{QCD considerations}
To extend this formalism to QCD, we introduce quark field\footnote{To maintain the definition of the trace, (fermionic) quark fields use anti-periodic {\em temporal} boundary conditions while (bosonic) gauge links use periodic {\em temporal} boundary conditions.} variables on the sites of the lattice, and connect them with gauge link variables representing the gluonic degrees of freedom.  The gauge links provide a connection between adjacent sites that maintains color gauge-covariance.
We use a finite temporal extent $\beta=N_\tau a$, and render the number of integrals on each time slice finite by working in a finite periodic spatial volume $L^3=N_s a^3$, where  $N_\tau$ and $N_s$ are the number of temporal and spatial lattice sites, respectively.
The temporal extent $\beta$ in Eq.~\ref{eqn:partition2} is analogous to the inverse temperature in statistical mechanics:
 \begin{equation}
 \beta = N_\tau a \sim 1/(\mbox{temperature}),
 \end{equation}
 and thus the lattice formalism can be used to calculate the QCD Equation of State and investigate such systems as the quark-gluon plasma~\cite{bib:thermo}.  To determine the QCD spectrum, we are interested in the zero-temperature limit of the partition function, and will work with large temporal extents $\beta\to\infty$.  It should be noted that Minkowski time does not appear anywhere in this formulation.  This does not cause difficulties because a theory's spectrum is an equilibrium (i.e. time-independent) property.  It is possible, however, to define lattice formulations which allow the investigation of non-equilibrium properties of QCD such as transport coefficients~\cite{bib:noneq}.

\subsection{The spectral representation of correlation functions}
To access the spectrum, we define a {\em correlator} $C^{(\beta)}(\tau)$ for $0<\tau<\beta$ by considering correlation between time-ordered source and sink operators defined on time slices separated\footnote{The temporal boundary conditions may be used to shift the source operator to time slice $0$ and the sink operator to time slice $\tau$.} by $\tau$ (c.f. Fig.~\ref{fig:lattice} and Eqs.~\ref{eqn:partition1}-\ref{eqn:partition2}):
\begin{equation}
C^{(\beta)}(\tau)\equiv\langle \mathcal{O}(\tau)\overline{\mathcal{O}}(0)\rangle \equiv \frac{1}{Z}\Tr{e^{- (\beta-\tau) H}\mathcal{O}e^{-\tau H}\overline{\mathcal{O}}},\quad Z=\Tr{e^{-\beta H}}.\label{eqn:correlator}
\end{equation}

We may express the traces in terms of the energy eigenstates: 
\begin{eqnarray}
C^{(\beta)}(\tau) &=& \frac{1}{Z}\sum_{n=0}^\infty \bra{n}e^{- (\beta-\tau) H}\mathcal{O}e^{-\tau H}\overline{\mathcal{O}}\ket{n}, \quad H\ket{n}=E_n\ket{n}, \quad E_0 = 0\\
&=&\frac{1}{Z}\sum_{n=0}^\infty e^{-(\beta-\tau) E_n} \bra{n}\mathcal{O}e^{-\tau H}\overline{\mathcal{O}}\ket{n},\quad Z=1+\sum_{n=1}^\infty e^{-\beta E_n}.
\end{eqnarray}
Finally, we take $\beta >> \tau$ to access the zero-temperature physics:
\begin{equation}
C^{(\beta)}(\tau)\stackrel{\beta >> \tau}{\to}\bra{0}\mathcal{O}e^{-\tau H}\overline{\mathcal{O}}\ket{0}\equiv C(\tau).\label{eqn:corr}
\end{equation}
We may decompose the zero-temperature correlator $C(\tau)$ defined in Eq.~\ref{eqn:corr} into its spectral components by inserting a complete set of energy states $\ket{k}$:
\begin{eqnarray}
C(\tau) &=& \bra{0}\mathcal{O}e^{-H\tau}\overline{\mathcal{O}}\ket{0},\\
&=&\bra{0}\mathcal{O}\sum_{k=0}^\infty \ket{k}\bra{k}e^{-H\tau}\overline{\mathcal{O}}\ket{0},\qquad H\ket{k}=E_k\ket{k},\\
&=&\sum_{k=1}^\infty |\bra{k} \overline{\mathcal{O}}\ket{0}|^2 e^{-E_k \tau},\qquad \mbox{choosing }\bra{0}\overline{\mathcal{O}}\ket{0}=0.\label{eqn:spectral_decomp}
\end{eqnarray}
The derivation of Eq.~\ref{eqn:spectral_decomp} neglects boundary conditions, and is therefore accurate only in the infinite-volume limit.
In this limit we see that $C(\tau)$ decays as a sum of exponentials with decay constants given by the energy levels accessible by the application of $\overline{\mathcal{O}}$ to the vacuum state of the system.  For large Euclidean time separations $\tau$, we see that the correlator is dominated by the energy gap between the vacuum and the first excited state.   It is crucial to design operators having $|\braket{k|\overline{\mathcal{O}}|0}|^2$ large for the states $\ket{k}$ of interest, and small for the other (contaminating) modes.  An overview of the LHPC's operator construction method can be found elsewhere in these proceedings~\cite{bib:bulava}, and more detailed descriptions may be found in~\cite{bib:colin_gt, bib:lichtl_dissertation}.

\section{Improving signal quality}
In practice, the correlator $C(\tau)$ is estimated using the Monte Carlo method~\cite{bib:colin_mc}, and will consequently have an associated uncertainty for each value of $\tau$.  It can be shown~\cite{bib:lepage} that the signal-to-noise ratio for baryon correlators decays exponentially with $\tau$.
Recalling the expression for the correlator given in Eq.~\ref{eqn:spectral_decomp}:
$$C(\tau)=\sum_{k=1}^\infty |\bra{k} \overline{\mathcal{O}}\ket{0}|^2 e^{-E_k \tau}$$
we see that if the operator $\overline{\mathcal{O}}$ couples strongly to several states $\ket{k}$, the first excited energy level $E_1$ may not have a chance to dominate before the signal is lost in the noise.

\begin{figure}
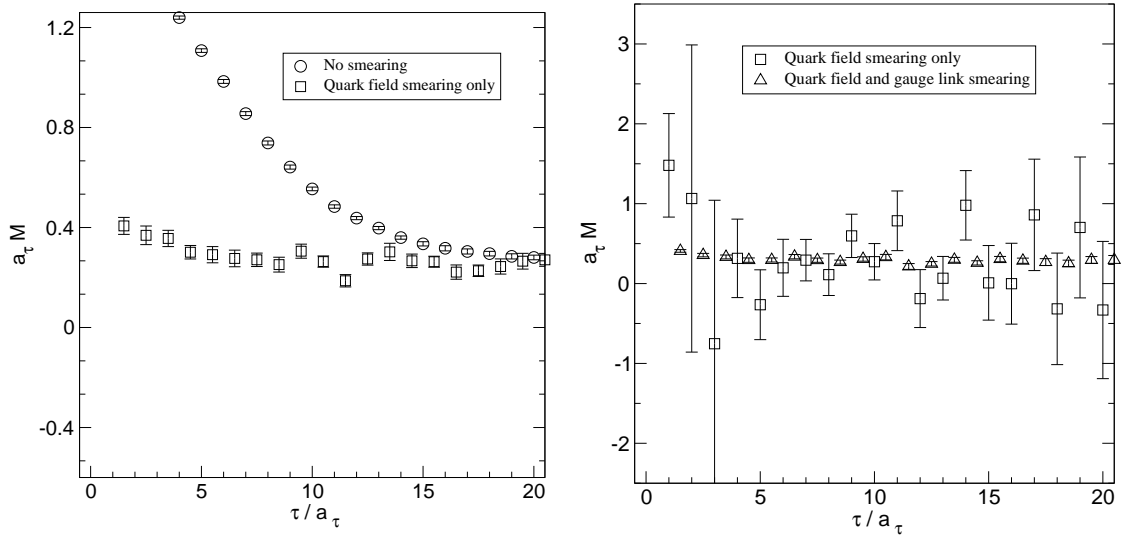
 
\begin{minipage}{0.49\textwidth}
\includegraphics[width=\textwidth]{meff-contaminated_good}
\end{minipage}
\hspace{0.01\textwidth}
\begin{minipage}{0.49\textwidth}
\includegraphics[width=\textwidth]{meff-noisy_good}
\end{minipage}
\caption{Left: a local operator's effective mass signal.  Quark field smearing drastically reduces the operator's coupling to short-wavelength contaminating modes. Right: an extended operator's effective mass signal (after quark field smearing).  Gauge link smearing strongly attenuates the noise coming from the stochastic evaluations of the gauge links used in constructing the operator.\label{fig:meff}}
\end{figure}

 We may define the {\em effective mass function}, which becomes the lowest energy (in lattice units) at large values of $\tau$:
\begin{equation}
a M(\tau) = \ln\left[C(\tau)/C(\tau+a)\right]\stackrel{\tau\to\infty}{\to}a E_1.
\end{equation}
Effective mass plots, such as those shown in Fig.~\ref{fig:meff}, are a useful tool for visualizing and evaluating an operator's correlator signal according to two key criteria: (1) excited state contamination, and (2) noise.

Fig.~\ref{fig:meff} demonstrates two powerful ways to improve signal quality: quark field smearing and gauge link smearing~\cite{bib:lichtl_smear}.  The smearing process replaces the variables at each location with a suitable form of a weighted local spatial average.  It has been found that quark field smearing drastically reduces the coupling of operators to the short-wavelength contaminating modes of the theory, at the price of a modest increase in noise.  The primary source of noise in an operator, especially in {\em extended} operators using quarks which are covariantly displaced from one another, is the presence of stochastically updated gauge link variables.  The application of link smearing strongly attenuates the noise of the operator.

\section{Extracting excited resonances}
\subsection{Using the variational method to extract excited states}

\begin{figure}[b]
\begin{minipage}{0.49\textwidth}
\includegraphics[width=\textwidth]{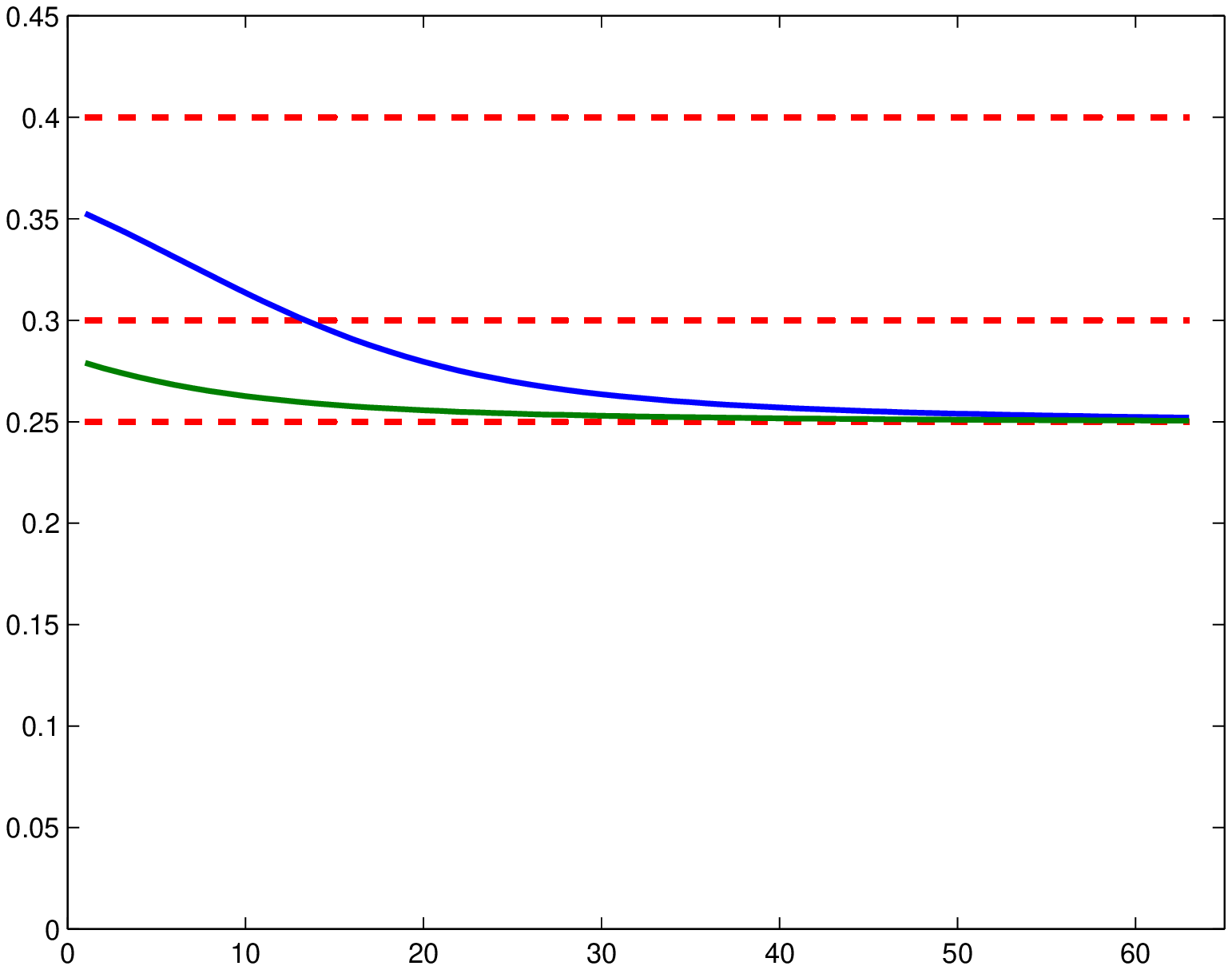}
\end{minipage}
\hspace{0.01\textwidth}
\begin{minipage}{0.49\textwidth}
\includegraphics[width=\textwidth]{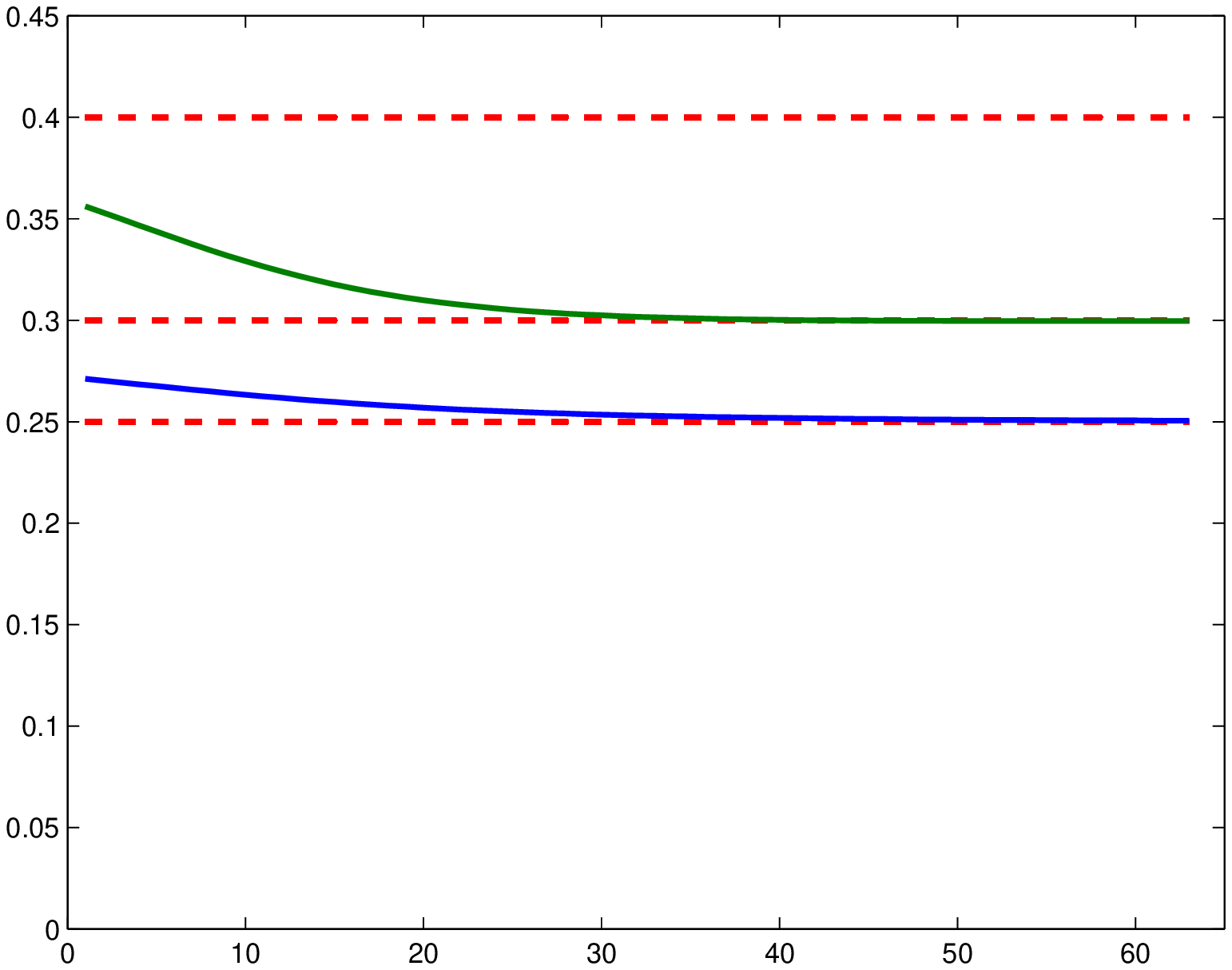}
\end{minipage}
\caption{A toy theory with three energy levels shown as dashed horizonal lines.  Two operators are used to form a $2\times2$ correlator matrix, and the effective masses associated with the diagonal elements $C_{aa}(\tau)$ are shown on the left.  Diagonalizing the correlator matrix yields the principal correlators $v^\dagger_iC(\tau)v_i$, whose effective masses are shown on the right.  This method can be used to extract multiple excited states.\label{fig:toy_model}}
\end{figure}

Instead of one operator $\overline{\mathcal{O}}$, we may use a {\em basis} of operators $\{\overline{\mathcal{O}}_a\}$ to define a {\em correlator matrix}:
\begin{equation}
\langle \mathcal{O}_a(\tau)\overline{\mathcal{O}}_b(0)\rangle\stackrel{\beta>>\tau}{\to}\bra{0}\mathcal{O}_a e^{-H \tau}\overline{\mathcal{O}}_b\ket{0}\equiv C_{ab}(\tau).
\end{equation}
We may view the quantities $C_{ab}(\tau)$ as matrix elements between states in the {\em trial basis}$\\\ket{\mathcal{O}_a}\equiv \overline{\mathcal{O}}_a\ket{0}$.
Once we have estimates of these elements, we may apply the variational method~\cite{bib:prineff} to define a new operator basis
$\overline{\Theta}_i \equiv \sum_a\overline{\mathcal{O}}_a v_{ai}$,
and may choose the coefficient vectors $v_i$ to diagonalize $e^{-H\tau}$ {\em in the subspace spanned by our trial basis}:
\begin{eqnarray}
v_i^\dagger C(\tau)v_j&=&\bra{0}\Theta_i\left[e^{-H\tau}\right]\overline{\Theta}_j\ket{0},\\
&=&0\mbox{ if } i\neq j.
\end{eqnarray}

This orthogonality implies that the diagonal matrix elements $v^\dagger_iC(\tau)v_i$, the so-called {\em principal correlators}, will be asymptotically dominated by different energy levels, as shown in Fig.~\ref{fig:toy_model}.  One refinement of the method is to relax the trial basis by operating upon it with $\exp(-H\tau_0)$ for some small relaxation time $\tau_0$.  Thus, by diagonalizing 
\begin{equation}
C^{-1/2}(\tau_0)C(\tau)C^{-1/2}(\tau_0),
\end{equation}
it can be shown~\cite{bib:lichtl_lat07} that one is working (formally) with the basis of trial states:
\begin{equation}
\ket{\mathcal{O}_a(\tau_0)}\equiv \exp(-H\tau_0/2)\overline{\mathcal{O}}_a\ket{0}.
\end{equation}
By relaxing the trial basis, we obtain a better overlap with the subspace spanned by the low-lying states of interest.  In practice, this method is complicated by the presence of noise in the estimates of $C_{ab}(\tau)$.

\clearpage

\begin{figure}[t]
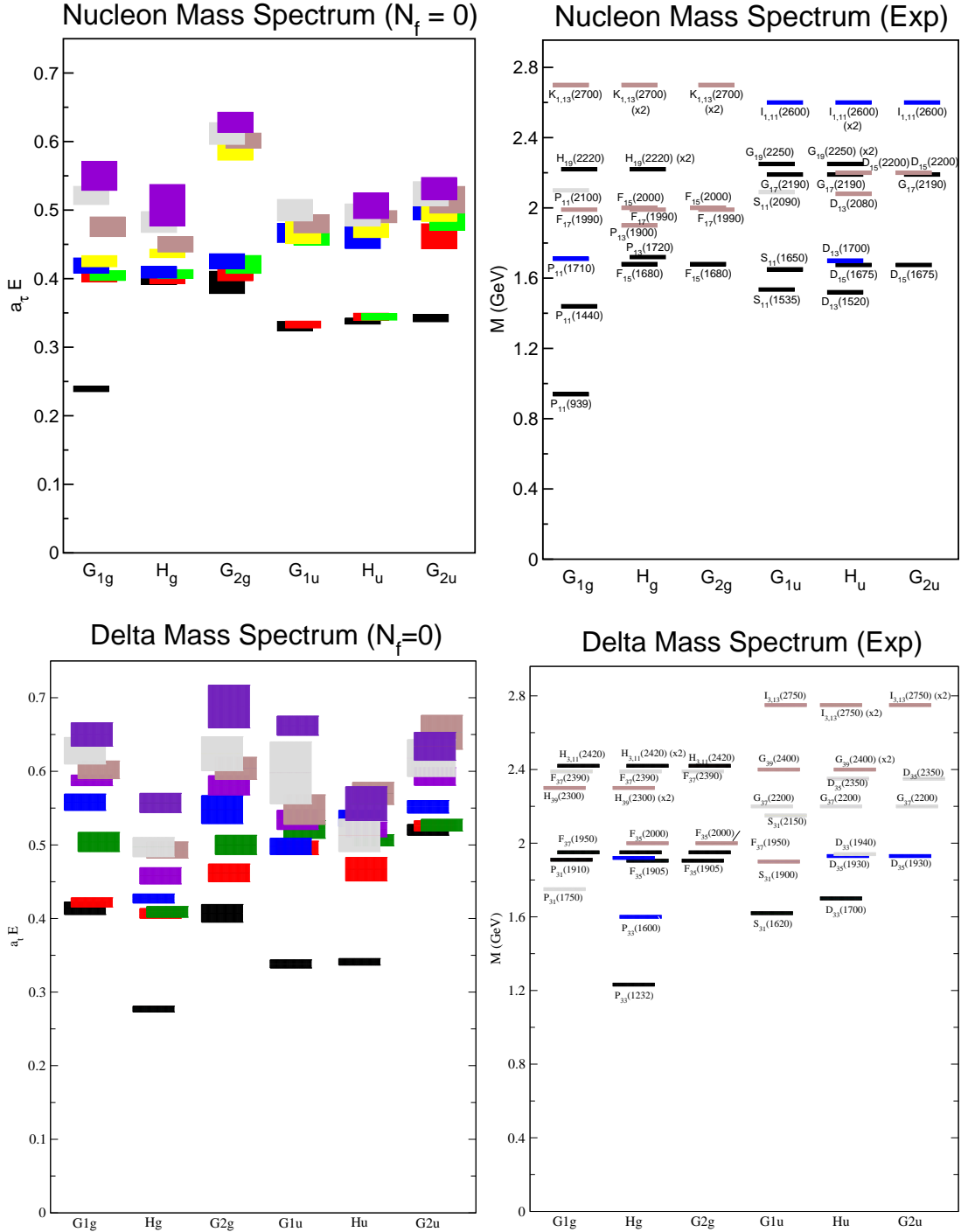

\begin{minipage}{.49\textwidth}
\includegraphics[width=0.99\textwidth]{nucleon_spectrum_nf0}\\
\phantom{0}\\
\includegraphics[width=0.99\textwidth]{DeltaMasses}
\end{minipage}
\begin{minipage}{.49\textwidth}
\includegraphics[width=0.99\textwidth]{nucleon_spectrum_expt}\\
\phantom{0}\\
\includegraphics[width=0.99\textwidth]{DeltaMasses_Expmnt}
\end{minipage}
\caption{
Left: The low-lying Nucleon (top) and Delta (bottom) baryon spectra as determined by 200 quenched (``zero flavors of dynamical quarks'') configurations on a $12^3\times 48$ anisotropic lattice with $a_s\sim 0.1$ fm and $a_s/a_\tau\sim 3.0$.  The quenched approximation neglects the effects of dynamical sea quarks, and corresponds to neglecting quark loops in diagrammatic calculations.  Our pion mass for this exploratory study was approximately 700 MeV. The vertical height of each box indicates the statistical
uncertainty in that estimate.
Right:  The corresponding spectra as determined by
experiment~\cite{PDG} and projected onto the space of lattice spin-parity
 states.  While there are infinitely many continuum $J^p$ quantum numbers, there are only a finite number of irreducible representations of the spinorial lattice rotation group $O_h^D$: $G_{1g}$, $H_g$, and $G_{2g}$ having even parity, and $G_{1u}$, $H_u$, $G_{2u}$ having odd parity.  Consequently, continuum $J^p$ states will appear on the lattice as degenerate levels in one or more lattice spin-parity channels.  For example, the continuum $D_{15}(1675)$ $J^p=\frac{5}{2}^+$ resonance appears on the lattice as a pair of degenerate levels appearing in the $H_u$ and $G_{2u}$ spin-parity channels.
\label{fig:spectra}}
\end{figure}

\clearpage

\section{Results and outlook}

Lattice QCD calculations of the Nucleon and Delta baryon spectra~\cite{bib:bulava, bib:lichtl_dissertation} are presented in Fig.~\ref{fig:spectra}.  Comparison is made with experimental results~\cite{PDG} by subducing continuum $J^p$ quantum numbers  onto their discrete lattice spin-parity counterparts~\cite{bib:colin_gt} (even parity: $G_{1g}$, $H_g,$ $G_{2g}$, odd parity: $G_{1u}$, $H_u$, $G_{2u}$). Although this exploratory study is {\em quenched} (neglecting the effects of dynamical sea quarks) and performed at an unphysically high pion mass of approximately 700 MeV, several interesting patterns are seen.

The quenched spectra reproduce the isolated even-parity states corresponding to the proton and the $\Delta^{++}(1232)$.  We also see a band of low-lying odd-parity states in each spectrum matching the corresponding low-lying experimental odd-parity bands.  In contrast to experiment, we find that the first excited even-parity states corresponding to the Roper and the $P_{33}(1600)$ resonances lie {\em above} the first odd-parity band of states. Also, the splitting between the first excited even-parity band and the lowest-lying odd-parity band of states is significantly larger in the quenched spectrum.  These discrepancies may be due to quenching, sensitivity to the quark mass, finite-volume effects, discretization artifacts, or poor interpolation by our three-quark operators.
Unquenched runs currently underway at multiple volumes and pion masses will attempt to resolve these issues.

In conclusion, the lattice formulation is a powerful method for calculating non-perturbative quantities used in fundamental tests of QCD.  The exploratory results presented here demonstrate clear progress toward the long-standing goal of determining the QCD resonance spectrum from first principles.  Ongoing studies using improved operators, variational methods, and dynamical sea quarks will continue to reveal the patterns present in the QCD spectrum.

This research is supported by NSF grant PHY 0653315, and numerical calculations were performed using the Chroma QCD library~\cite{Chroma} on the Carnegie Mellon University Medium Energy Group computing cluster.  Additional resources and travel support to the VII Latin American Symposium on Nuclear Physics and Applications were provided by the RIKEN BNL Research Center.

\end{document}